\documentclass{PoS}

\title{Scaling properties of the chiral phase transition in the low density region of two-flavor QCD with improved Wilson fermions}

\ShortTitle{Scaling properties of the chiral phase transition in the low density region of two-flavor QCD with ...}

\author{\speaker{T.~Umeda}$^1$, S. Ejiri$^2$, K.~Kanaya$^3$, Y.~Maezawa$^4$, 
Y.~Nakagawa$^2$, H.~Ohno$^5$, H.~Saito$^6$, S.~Yoshida$^{5,7}$ (WHOT-QCD Collaboration)  \\ \\
$^1$Graduate School of Education, Hiroshima University, Hiroshima 739-8524, Japan \\
      \ \ \ E-mail: \email{tumeda@hiroshima-u.ac.jp} \\
$^2$Graduate School of Science and Technology, Niigata University, Niigata 950-2181, Japan \\
$^3$Graduate School of Pure and Applied Sciences, University of Tsukuba, 
Tsukuba, Ibaraki 305-8571, Japan \\
$^4$Fakult\"at f\"ur Physik, Universit\"at Bielefeld, D-33615 Bielefeld, Germany \\
$^5$Physics Department, Brookhaven National Laboratory, Upton, NY 11973, USA \\
$^6$NIC, DESY Zeuthen, Platanenallee 6, 15738 Zeuthen, Germany \\
$^7$Theoretical Research Division, Nishina Center, RIKEN, Wako 351-0198, Japan 
}

\abstract{
We study scaling behavior of a chiral order parameter in the low density region, performing a simulation of two-flavor QCD with improved Wilson quarks. The scaling behavior of the chiral order parameter defined by a Ward-Takahashi identity agrees with the scaling function of the three-dimensional O(4) spin model at zero chemical potential. We extend the scaling study to finite density QCD. Applying the reweighting method and calculating derivatives of the chiral order parameter with respect to the chemical potential, the scaling properties of the chiral phase transition are discussed in the low density region. We moreover calculate the curvature of the phase boundary of the chiral phase transition in the temperature and chemical potential plane assuming the O(4) scaling relation.}

\FullConference{31st International Symposium on Lattice Field Theory - LATTICE 2013\\
		July 29 - August 3, 2013\\
		Mainz, Germany}

\begin{document}

\section{Introduction}

\begin{figure}[tb]
\begin{minipage}{7.3cm}
\begin{center}
\includegraphics[width=63mm]{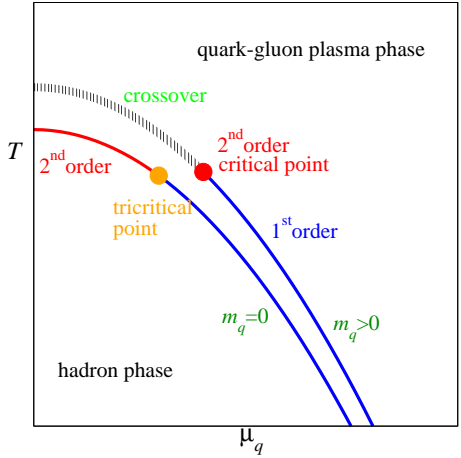}
\vskip -0.3cm
\caption{
Speculated phase structure of 2-flavor QCD at finite temperature and density. The red line is the second order transition line for $m_q=0$ and the blue lines are the first order transition lines.
}
\label{fig1}
\end{center}
\end{minipage}
\hspace{3mm}
\begin{minipage}{7.3cm}
\begin{center}
\makeatletter
\def\@captype{table}
\makeatother
%{\renewcommand{\arraystretch}{1.0} \tabcolsep = 1mm
 \small
 \begin{tabular}{|ccccc|}
 \cline{1-5}
 \multicolumn{1}{|c}{$\beta$} &
 \multicolumn{1}{c} {$K$}    & 
 \multicolumn{1}{c} {$T/T_{pc}$} & 
 \multicolumn{1}{c}{Conf.} &
 \multicolumn{1}{c|}{$m_{\rm PS}/m_{\rm V}$} \\
 \cline{1-5}
 1.50 & 0.150290 & 0.82(3) & 2500 & 0.678(2)\\
 1.60 & 0.150030 & 0.86(3) & 2500 & 0.663(2) \\
 1.70 & 0.148086 & 0.94(3) & 2500 & 0.659(2) \\
 1.75 & 0.146763 & 1.00(4) & 2500 & 0.662(3) \\
 1.80 & 0.145127 & 1.07(4) & 2500 & 0.657(5) \\
 1.85 & 0.143502 & 1.18(4) & 500  & 0.652(4) \\
 1.90 & 0.141849 & 1.32(5) & 500  & 0.648(4) \\
 1.95 & 0.140472 & 1.48(5) & 500  & 0.657(4) \\
 \cline{1-5}
 1.50 & 0.143480 & 0.76(4) & 2550 & 0.820(1) \\
 1.60 & 0.143749 & 0.80(4) & 2600 & 0.809(1) \\
 1.70 & 0.142871 & 0.84(4) & 2600 & 0.804(1) \\
 1.80 & 0.141139 & 0.93(5) & 2600 & 0.800(2) \\
 1.85 & 0.140070 & 0.99(5) & 600  & 0.794(2) \\
 1.90 & 0.138817 & 1.08(5) & 600  & 0.796(2) \\
 1.95 & 0.137716 & 1.20(6) & 600  & 0.802(2) \\
 \cline{1-5}
\end{tabular}
\vspace{-1mm}
\caption{Simulation parameters.}
% for $m_{\rm PS}/m_{\rm V} \approx 0.65$ (left) 
% and $m_{\rm PS}/m_{\rm V} \approx 0.80$ (right).}
\label{tab1}
%}
\end{center}
\end{minipage}
\end{figure}

Many interesting properties of finite temperature and density QCD have been uncovered by lattice simulations. 
However, there are still many open problems even at low density. 
The nature of the chiral phase transition in the chiral limit of 2-flavor QCD is one of them.
The standard expectation, assuming the U$_{\rm A}$(1) symmetry remains violated also in the high temperature phase, is that the chiral phase transition in 2-flavor QCD is of second order in the chiral limit $m_q=0$ and crossover for $m_q \neq 0$, and it changes to first order when the chemical potential $\mu_q$ is sufficiently large. 
In this case, the scaling property around the second order transition is universal to that of the 3-dimensional O(4) spin model.
We illustrate the conjecture in Fig.~\ref{fig1} for 2-flavor QCD with $m_q=0$ and $m_q \neq 0$. 
Because the QCD action has the chiral symmetry in the chiral limit even at $\mu_q\ne0$, 
we expect the same critical properties in the low density region \cite{bnl-bie10}. 

The O(4) scaling behavior in QCD was first reported for the case of Wilson-type quark actions at $\mu_q=0$.
Both with the standard Wilson quark action \cite{iwasaki} and with the clover-improved Wilson quark action \cite{cppacs1}, 
a subtracted chiral condensate is shown to follow the scaling behavior with the critical exponents and scaling function of the O(4) spin model in a rather wide range of the parameter space. 
Studies using improved staggered quark actions have also shown that, adopting several definitions for the renormalized chiral condensate, the chiral scaling is consistent with O(4) and O(2) with very light u,d quark masses and the physical strange mass \cite{bnl-bie09} although the universality is not guaranteed due to the explicit violation of locality due to the forth-root trick. 
On the other hand, it was recently argued that the U$_{\rm A}$(1) symmetry may be effectively recovered in the high temperature phase \cite{aoki12}. 
This suggests that the chiral condensate does not follow the O(4) scaling. 
Hence, it is worth revisiting the scaling study both at zero and finite densities.

In this report, we study the scaling behavior near the chiral phase transition in the low density region of 2-flavor QCD. 
To avoid theoretical uncertainties with the forth-root trick, we adopt improved Wilson quarks. 
General argument of the O(4) scaling at $\mu_q \neq 0$ is given in Sec.~\ref{sec:scaling}.  
Numerical results of the scaling tests are presented in Sec.~\ref{sec:result} and
%  we then discuss the $\mu_q$ dependence of the critical line in Sec.~
\ref{sec:curve}. 
A conclusion is given in Sec.~\ref{sec:summary}.

\section{Scaling behavior of chiral order parameter at finite density}
\label{sec:scaling}

The order parameter in the O(4) spin model is given by the magnetization $M$.
In the vicinity of the second order transition point, $M$ satisfies the following scaling relation:
\begin{eqnarray}
M/h^{1/\delta} = f(t/h^{1/y}),
\label{eq:o4}
\end{eqnarray}
where $h$ is the external magnetic field, $t$ is the reduced temperature, $t=(T-T_c|_{h=0})/T_c|_{h=0}$,  
and $f(x)$ is the scaling function.
In the O(4) spin model, the critical exponents are 
$1/y \equiv 1/(\beta \delta) = 0.546$ and $1/ \delta = 0.2073(4)$ \cite{engels00}. 
In 2-flavor QCD, the scaling variables, $M$, $t$ and $h$, may be identified \cite{bnl-bie10} as 
\begin{eqnarray}
M = \langle \bar{\psi} \psi \rangle,
\hspace{5mm}
t= \beta - \beta_{ct} + \frac{c}{2} \left(\frac{\mu_q}{T} \right)^2, 
\hspace{5mm}
h=2m_q a, 
\label{eq:sclv}
\end{eqnarray}
respectively, where $\beta_{ct}$ is the critical point of $\beta=6/g^2$ at $\mu_q=0$ in the chiral limit, 
$a$ is the lattice spacing, 
and $c$ is the curvature of the critical line in the $(\beta, \mu_q/T)$ plane,
$c \equiv - d^2 \beta_{ct}/d(\mu_q/T)^2$, since 
$\beta_{ct}(\mu_q) = \beta_{ct}(0) - c(\mu_q/T)^2 /2$ on the critical curve $(t=0)$.
Here, $h$ does not have a $\mu_q$-dependent term at low density. 
Because the critical line is expected to run along the $m_q=0$ axis in the low density region of the $(m_q, \mu_q/T)$ plane, $h=0$ at $m_q=0$.

We compare the scaling functions of 2-flavor QCD and the O(4) spin model in the vicinity of $\mu_q=0$. Although a direct simulation of lattice QCD is difficult at $\mu_q \ne 0$ due to the complex weight problem, the reweighting method is applicable at small $\mu_q$.
In this note, we study the scaling property of the second derivative of the chiral order parameter, 
\begin{eqnarray}
\left. \frac{d^2 M}{d(\mu_q/T)^2} \right|_{\mu_q=0} 
= c \left. \frac{dM}{dt} \right|_{\mu_q=0}, \hspace{5mm}
\left. \frac{dM/dt}{h^{1/\delta-1/y}} \right|_{\mu_q=0} 
= \left. \frac{df(x)}{dx} \right|_{x=t/h^{1/y}}.
\label{eq:der2sc}
\end{eqnarray}
Assuming these scaling relations, the coefficient $c$ 
%in the scaling variable Eq.~(\ref{eq:sclv}) 
corresponds to the curvature of the critical line at $\mu_q=0$ in the chiral limit. 
%It may be worth discussing the phase boundary by measuring $c$.

Here, a careful treatment is required because the chiral symmetry is explicitly broken with Wilson quarks at finite $a$. 
In Ref.~\cite{iwasaki,cppacs1}, it was shown that the O(4) scaling of Eq.~(\ref{eq:o4}) is well satisfied when one defines the quark mass $m_q a$ and the chiral order parameter $\langle \bar{\psi} \psi \rangle$ by Ward-Takahashi identities \cite{bochicchio}.
%The quark mass 
$m_q a$ can be defined by 
%for temporal direction:
\begin{eqnarray}
2m_q a = -m_{\rm PS} \left. \langle \bar{A_4}(t) \bar{P}(0) \rangle \right/
\langle \bar{P}(t) \bar{P}(0) \rangle,
\end{eqnarray}
where $P$ and $A_{\mu}$ are the pseudo-scalar and axial-vector meson operators, respectively, 
$m_{\rm PS}$ is the pseudo-scalar meson mass, and the bar means the spatial average.
Similarly, $\langle \bar{\psi} \psi \rangle$ is given by
\begin{eqnarray}
\langle \bar{\psi} \psi \rangle
= \frac{2 m_q a}{N_s^3 N_t} \sum_{x,x'} \langle P(x) P(x') \rangle
= \frac{2 m_q a (2K)^2 }{N_s^3 N_t} \left\langle 
{\rm tr} \left( D^{-1} \gamma_5 D^{-1} \gamma_5 \right) \right\rangle.
\end{eqnarray}
Here, $D$ is the quark matrix, $K$ is the hopping parameter, and $N_s^3 \times N_t$ is the number of sites. 
The quark mass and the chiral condensate satisfy the Ward-Takahashi identity in the continuum limit:
%\begin{eqnarray}
$
\langle \partial_{\mu} A_{\mu} (x) P(x') \rangle
-2m_q a \langle P(x) P(x') \rangle
=\delta(x- x') \langle \bar{\psi} \psi \rangle.
$
%\end{eqnarray}
We adopt these definitions of quark mass and chiral order parameter in this study.

\section{Chiral order parameter at finite density}
\label{sec:result}

We perform simulations of 2-flavor QCD at finite temperature and $\mu_q=0$ on a $16^3 \times 4$ lattice and combined them with configurations obtained in Refs.~\cite{whot07,whot09}. 
The RG-improved gauge action and the 2-flavor clover-improved Wilson quark action are adopted. The measurements are done every 10 trajectories and 500 -- 2600 configurations are used for the analysis at each simulation point. 
The simulation parameters are summarized in Table \ref{tab1}.
%The details of the simulation are given in Ref.~\cite{whot07,whot09}.
The quark mass $m_q a$ is computed performing zero temperature simulations 
on a $16^3 \times 24$ lattice at each simulation point listed in Table \ref{tab1}.
The number of configurations used for the measurement is 378 -- 589. 
The pseudo-scalar to vector meson mass ratio at $T=0$ is about 
$m_{\rm PS}/m_{\rm V} \approx 0.65$ or $0.80$, as shown in Table \ref{tab1}.

We use the random noise method to calculate $\langle \bar{\psi} \psi \rangle$. 
As we have emphasized in Ref.~\cite{whot09}, it is important to apply the noise method only for the space index and to solve the inverse exactly for the spin and color indices without applying the noise method to obtain reliable results.
We choose 100 -- 150 as the number of noise vectors for each color and spin indices.

\subsection{Reweighting method for the chiral order parameter}
\label{sec:rew}

We use the reweighting method to calculate $\langle \bar{\psi} \psi \rangle$
at finite $\mu_q$,
\begin{eqnarray}
(2K)^2 \left\langle {\rm tr} \left( D^{-1} \gamma_5 D^{-1} \gamma_5 \right) 
\right\rangle_{\beta, \mu_q}
&=& (2K)^2 \frac{1}{Z} \int {\cal D} U \ {\rm tr}(D^{-1} \gamma_5 D^{-1} \gamma_5 ) 
(\det D)^{N_{\rm f}} e^{-S_g} \nonumber \\
&=& \frac{ (2K)^2 \left\langle 
{\rm tr} \left( D^{-1} \gamma_5 D^{-1} \gamma_5 \right)(\mu_q) 
e^{N_{\rm f} ( \ln \det D(\mu_q) - \ln \det D(0))} \right\rangle_{\beta, 0}
}{\left\langle e^{N_{\rm f} ( \ln \det D(\mu_q) - \ln \det D(0))} 
\right\rangle_{\beta, 0}}, \ \ 
\label{eq:rewchi}
\end{eqnarray}
where $N_{\rm f}=2$.
Because the reweighting method is applicable only for small $\mu_q$, 
we evaluate $\ln \det M(\mu_q)$ and 
${\rm tr} ( D^{-1} \gamma_5 D^{-1} \gamma_5)$ 
by a Taylor expansion up to $O(\mu_q^2)$, as proposed on Ref.~\cite{BS02}, 
\begin{eqnarray}
N_{\rm f} (\ln \det D(\mu_q) - \ln \det D(0)) &=& \mu_q a {\cal Q}_1 
%\frac{d \ln \det D}{d \mu}
+ \frac{(\mu_q a)^2}{2} {\cal Q}_2 + O(\mu_q^3), \nonumber \\
%\frac{d^2 \ln \det D}{d \mu^2} +O(\mu^3),
(2K)^2 {\rm tr} \left( D^{-1} \gamma_5 D^{-1} \gamma_5 \right)(\mu_q) 
&=& (2K)^2 {\rm tr} \left( D^{-1} \gamma_5 D^{-1} \gamma_5 \right)(0) 
+ \mu_q a {\cal C}_1 
%\frac{d {\rm tr} \left( D^{-1} \gamma_5 D^{-1} \gamma_5 \right)}{d \mu}
+ \frac{(\mu_q a)^2}{2} {\cal C}_2 +O(\mu_q^3). \ \ \ \ \
%\frac{d^2 
%{\rm tr} \left( D^{-1} \gamma_5 D^{-1} \gamma_5 \right)}{d \mu^2}
\label{eq:dpipi}
\end{eqnarray}
where ${\cal Q}_n$ and ${\cal C}_n$ are defined by
\begin{eqnarray}
{\cal Q}_n = N_{\rm f} \frac{\partial^n \ln \det D} 
{\partial (\mu_q a)^n}, 
\hspace{5mm}
{\cal C}_n = (2K)^2 \frac{\partial^n {\rm tr} 
\left( D^{-1} \gamma_5 D^{-1} \gamma_5 \right)} {\partial (\mu_q a)^n}. 
\label{eq:basic}
\end{eqnarray}
These derivative operators can be calculated by the random noise method.

\begin{figure}[tb]
\begin{center}
\begin{minipage}{7.0cm}
\hspace{-7mm}
\includegraphics[width=8.6cm]{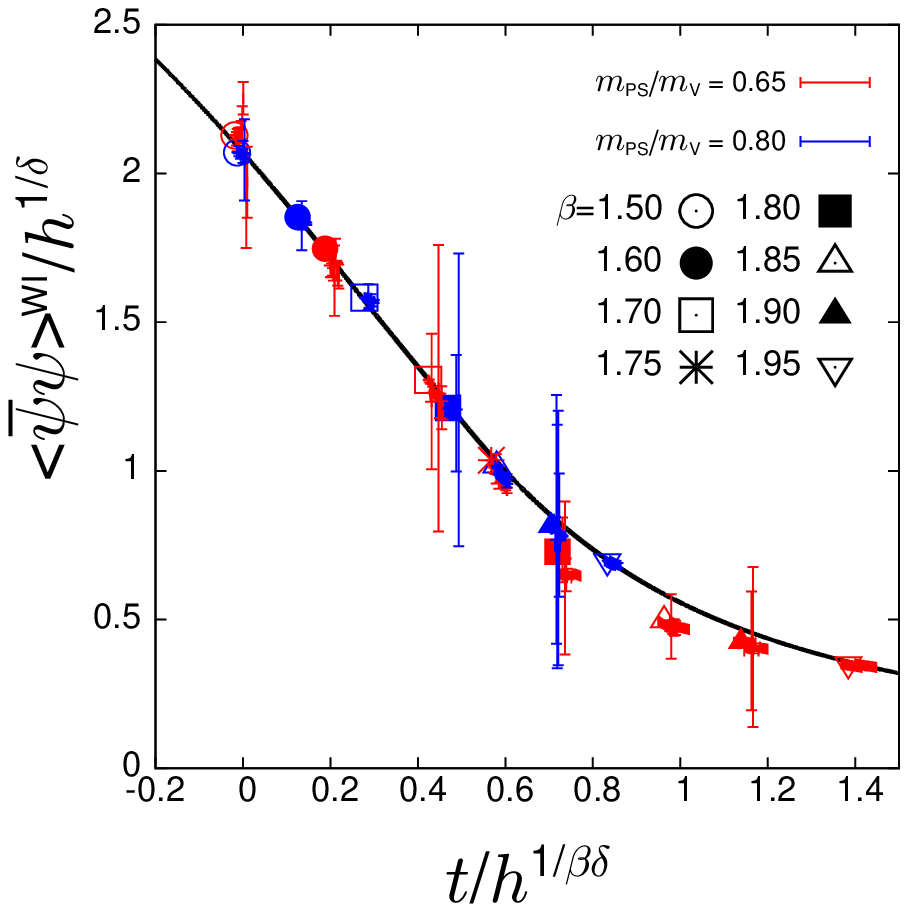}
\vskip -0.2cm
\caption{
O(4) scaling plot of the chiral order parameter in 2-flavor QCD. 
}
\label{fig2}
\end{minipage}
\hspace{2mm}
\begin{minipage}{7.0cm}
\hspace{-9mm}
\includegraphics[width=8.4cm]{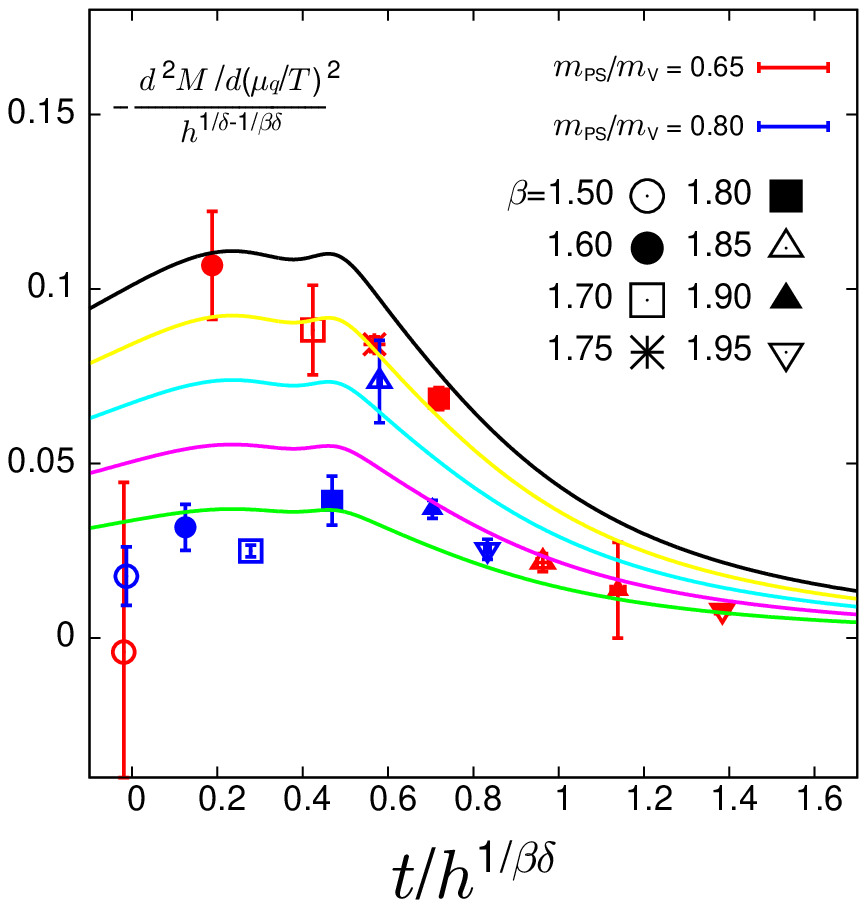}
\vskip -0.2cm
\caption{
O(4) scaling plot of the second derivative of the chiral order parameter by method 1.
}
\label{fig3}
\end{minipage}
\end{center}
\end{figure}

We plot the results of the chiral condensate in Fig.~\ref{fig2} for $\mu_q/T <1$,
constructing the form of $M/h^{1/\delta} = f(t/h^{1/y})$ with the identification of Eq.~(\ref{eq:sclv}).
The large symbols are the data at $\mu_q=0$. The results at finite $\mu_q$ are lines with their error bars connected with the sides of the symbols.
The critical exponents of the O(4) spin model are used. 
The black line is the scaling function obtained by the O(4) spin model in Ref.~\cite{engels00}. 
We adjust four fit parameters in this analysis. 
One is the critical value $\beta_{ct}$ at $\mu_q=0$, the second is the curvature of the phase boundary in the $(\beta, \mu_q/T)$ plane $c$, and the others are used for adjusting the scales of the horizontal and vertical-axes to the scaling function of the O(4) spin model.
We determine the four parameters such that the square of the deviation between the simulation data and the scaling curve is minimized. 
The best value of $\beta_{ct}$ and the curvature are $\beta_{ct}=1.510$ and $c=0.0290$.
This scaling plot indicates that the scaling function of 2-flavor QCD is consistent with that of the O(4) spin model, numerically.

\subsection{Derivatives of the chiral order parameter at $\mu_q=0$}
\label{sec:der}

We calculate the second derivative of $\langle \bar{\psi} \psi \rangle$ performing numerical 
simulations of 2-flavor QCD $(N_{\rm f}=2)$ at $\mu_q=0$ and 
compare it with the O(4) scaling function Eq.~(\ref{eq:der2sc}).
We then determine the curvature of the critical $\beta$ in the chiral limit, 
assuming the O(4) scaling behavior is satisfied.

\paragraph{Method 1: Fitting the data by the reweighting method at finite $\mu_q$}
We fit the data of the chiral order parameter at finite $\mu_q$ by
\begin{eqnarray}
\left\langle \bar{\psi} \psi \right\rangle (\mu_q)
= x + y (\mu_q/T)^2,
\label{eq:chifit}
\end{eqnarray}
where $x$ and $y$ are the fit parameters. 
The first derivative is zero due to the symmetry: $\mu_q \to -\mu_q$.
We identify the parameters as follows,
\begin{eqnarray}
x=\left\langle \bar{\psi} \psi \right\rangle (0), \hspace{5mm}
y= \frac{1}{2}
\frac{d^2 \left\langle \bar{\psi} \psi \right\rangle}{d (\mu_q/T)^2} (0).
\label{eq:fitpara}
\end{eqnarray}
We then obtain the second derivative and plot it in Fig.~\ref{fig3}
with the form of Eq.~(\ref{eq:der2sc}) adopting $\beta_{ct}=1.510$.
The corresponding scaling functions $df/dx$ are shown by the colored lines 
for $c=0.02, 0.03,0.04$ and $0.05$, from the bottom \cite{engels00}. 
The fit range is adopted to be $\mu_q/T <1$. 
The results change with the choice of fit range,
hence the systematic errors seem to be large in comparison to the statistic errors. 
Considering the size of systematic errors, clear deviation between the simulation result 
and the expected scaling function is not observed.

\begin{figure}[tb]
\begin{center}
\begin{minipage}{7.0cm}
\hspace{-9mm}
\includegraphics[width=8.4cm]{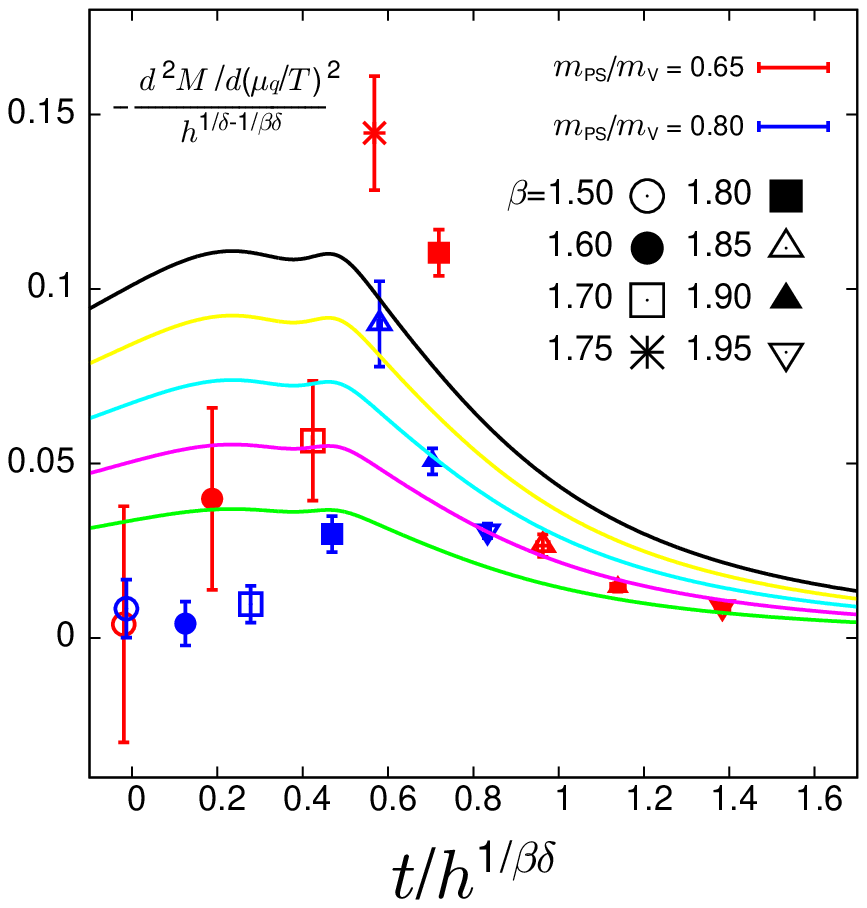}
\vskip -0.2cm
\caption{
O(4) scaling plot of the second derivative of the chiral order parameter by method 2.
}
\label{fig4}
\end{minipage}
\hspace{2mm}
\begin{minipage}{7.0cm}
\hspace{-9mm}
\includegraphics[width=8.4cm]{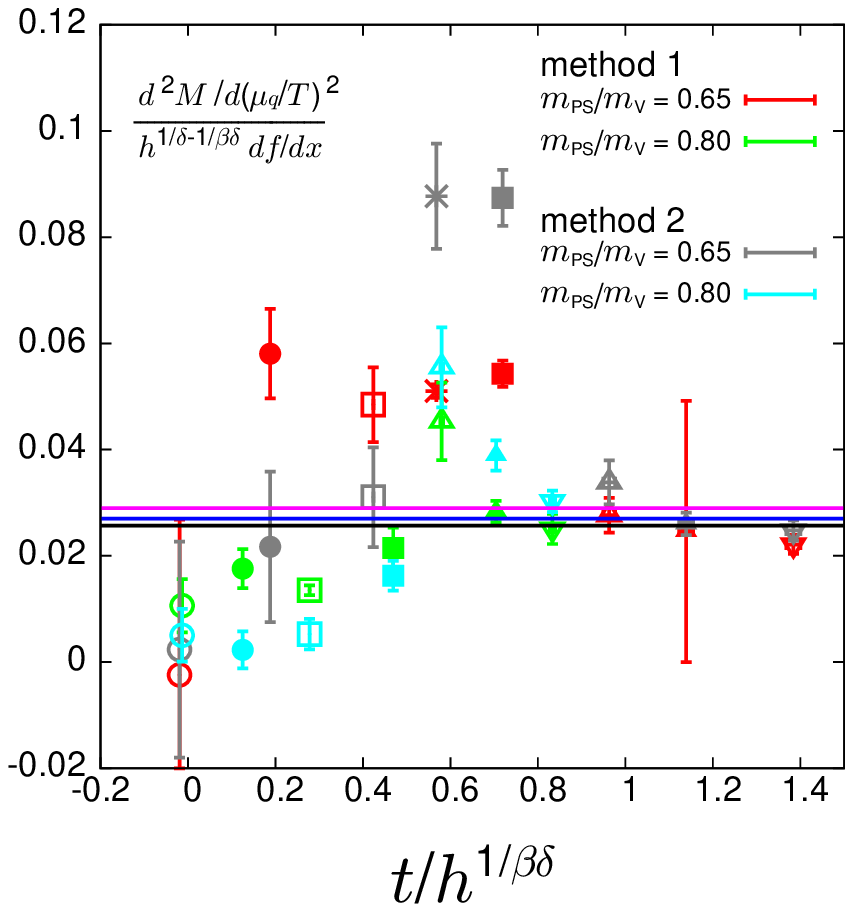}
\vskip -0.2cm
\caption{
The second derivative of the critical curve in the $(\beta, \mu_q/T)$ plane
$c$.
}
\label{fig5}
\end{minipage}
\end{center}
\end{figure}

\paragraph{Method 2: Computing the derivative operators}
Moreover, the derivative of the chiral order parameter is computed 
by the calculation of the following operators,
\begin{eqnarray}
&& {\cal A}_1 = \left\langle {\cal Q}_1 \right\rangle, \hspace{4mm}
{\cal A}_2 = \left\langle {\cal Q}_2 \right\rangle 
+\left\langle {\cal Q}_1^2 \right\rangle, \hspace{4mm}
{\cal F}_0 = 
\left\langle {\cal C}_0 \right\rangle, \nonumber \\ &&
{\cal F}_1 = \left\langle {\cal C}_1 \right\rangle 
+ \left\langle {\cal C}_0 {\cal Q}_1 \right\rangle, \hspace{4mm}
{\cal F}_2 = \left\langle {\cal C}_2 \right\rangle 
+ 2 \left\langle {\cal C}_1 {\cal Q}_1 \right\rangle 
+ \left\langle {\cal C}_0 {\cal Q}_2 \right\rangle 
+ \left\langle {\cal C}_0 {\cal Q}_1^2 \right\rangle.
\end{eqnarray}
Then, the derivatives of the chiral condensate are given by
\begin{eqnarray} 
\left\langle \bar{\psi} \psi \right\rangle
\biggr|_{\mu_q =0}
&=& \frac{2m_q a}{N_s^3 N_t} {\cal F}_0 , \hspace{8mm} 
\frac{\partial \left\langle \bar{\psi} \psi \right\rangle
}{\partial (\mu_q/T)}  \biggr|_{\mu_q =0} 
= 
%\frac{1}{N_t} \frac{\partial 
%\left\langle \bar{\psi} \psi \right\rangle}{\partial \mu} 
%\biggr|_{\mu=0}= 
\frac{2m_qa}{N_s^3 N_t^2} 
\left( {\cal F}_1 - {\cal F}_0 {\cal A}_1 \right) =0 , \nonumber \\
\frac{\partial^2 \left\langle \bar{\psi} \psi \right\rangle
}{\partial (\mu_q/T)^2}  \biggr|_{\mu_q=0} 
&=& 
%\frac{1}{N_t^2} \frac{\partial^2 
%\left\langle \bar{\psi} \psi \right\rangle}{\partial \mu^2} 
%\biggr|_{\mu=0}= 
\frac{2m_q a}{N_s^3 N_t^3} 
\left( {\cal F}_2 -2 {\cal F}_1 {\cal A}_1 - {\cal F}_0 {\cal A}_2 
+2 {\cal F}_0 {\cal A}_1^2 \right) 
= \frac{2m_q a}{N_s^3 N_t^3} 
\left( {\cal F}_2 - {\cal F}_0 {\cal A}_2 \right) ,
\end{eqnarray}
where we used the properties that
${\cal A}_n$ and ${\cal F}_n$ are zero for odd $n$'s at $\mu_q =0$.
%The operators, ${\cal C}_n, {\cal Q}_n$, can be calculated by a random noise method. 

The results of the second derivative are plotted in Fig.~\ref{fig4} 
with the form of Eq.~(\ref{eq:der2sc}).
The scaling functions $df/dx$ are also denoted similarly to Fig.~\ref{fig3}
%by the colored lines for $c=0.02, 0.03, 0.04$ and $0.05$, from the bottom. 
The difference between the results by method 1 and method 2 would be the systematic error, which is larger than the statistic error.
Although the uncontrollable systematical error is large, 
the simulation results roughly show the expected scaling behavior.

\section{Curvature of the critical line in the chiral limit}
\label{sec:curve}

Next, we estimate the second derivative of the critical $\beta$ with respect to $\mu_q$ in the chiral limit,
i.e.\ $c \equiv - d^2 \beta_{ct}/d(\mu_q/T)^2$, 
by three following methods, assuming the O(4) scaling behavior:
\begin{eqnarray}
\frac{1}{h^{1/\delta-1/y}}
\left. \frac{d^2 M}{d(\mu_q/T)^2} \right|_{\mu_q=0} 
= c \left. \frac{df(x)}{dx} \right|_{x=t/h^{1/y}}.
\label{eq:der2scfn}
\end{eqnarray}
Figure \ref{fig5} is a summary plot of the results of $c$.
The magenta line is the result by the global fit of the scaling plot in Sec. \ref{sec:rew}, 
i.e.\ $c = 0.0290$. 
The symbols are the ratio of $(d^2 M/d(\mu_q/T)^2) h^{-1/\delta +1/y}$ 
to $df(x)/dx$, which gives $c$.
The average of $c$ obtained by the reweighting method (method 1) 
is $c = 0.0273(42)$, shown by a blue line in Fig.~\ref{fig5}, and the average of the results from the second derivatives by the operator method (method 2) is 
$c = 0.0257(43)$, drawn by a black line.

The curvature of the critical temperature $T_c (\mu_q)$ at $\mu_q=0$ can be calculated by $c$ and the beta function as 
$(1/T_c) (d^2 T_c/d(\mu_q/T)^2) = c [a(d\beta/da)]^{-1}$.
%\begin{eqnarray}
%\frac{1}{T_c}
%\frac{d^2 T_c}{d(\mu_q/T)^2} 
%= - \left. \frac{d^2 \beta_{ct}}{d(\mu_q/T)^2} \right/ a \frac{d \beta}{da}. 
%\end{eqnarray}
This requires $a(d\beta/da)$ at $\beta_{ct}$ in the chiral limit.
The beta function $a(d\beta/da)$ for our lattice action was estimated in Ref.~\cite{cppacs2} for the pseudo-scalar-to-vector mass ratio $m_{PS}/m_{V} \geq 0.65$. 
Adopting $a(d\beta/da) \approx -0.5$ at $m_{PS}/m_{V} = 0.65$, we obtain
$-(1/T_c) (d^2 T_c/d(\mu_q/T)^2) \approx 0.05$--$0.06$.
This value is similar to that obtained using improved staggered quark actions in Refs.~\cite{bnl-bie10} and \cite{endrodi11}, 
but is much smaller than that of an experimental estimate for the chemical freeze out.

\section{Summary}
\label{sec:summary}

We discussed the scaling property of the chiral order parameter in the low density region of 2-flavor QCD.
The chiral order parameter and its second derivative were computed performing a simulation with improved Wilson-type quarks. We then compared the results with the O(4) scaling function.
The scaling behavior turned out to be roughly consistent with the O(4) universal scaling.
Assuming the O(4) scaling,
we estimated the curvature of the phase boundary in the $(\beta, \mu_q/T)$ plane.
However, to confirm the scaling property of 2-flavor QCD, 
a systematic study varying lattice volume and spacing is needed.

\paragraph{Acknowledgments}
%We would like to thank    for valuable discussions.
This work is in part supported by Grants-in-Aid of the Japanese 
%MEXT
Ministry of Education, Culture, Sports, Science and Technology 
(Nos.\ 21340049, 22740168, and 23540295) and 
by the Large Scale Simulation Program No.12/13-14 (FY2012-2013) of High Energy Accelerator 
Research Organization (KEK).
SY is supported by JSPS Strategic Young Researcher Overseas Visits Program
for Accelerating Brain Circulation (No.\ R2411).
HS is also supported by JSPS for Young Scientists.

\end{document}